\documentclass[12pt]{article}
\usepackage{verbatim,graphics,graphicx,color,slashed,amsmath}
\allowdisplaybreaks
\usepackage{bm}
\usepackage{cite}
\usepackage{amssymb}
\usepackage{braket}
\usepackage{ascmac}
\usepackage{multirow}
\usepackage[]{hyperref}
\hypersetup{colorlinks,bookmarks,unicode,linktocpage=true,linkcolor=blue, anchorcolor=blue, citecolor=blue}
\newcommand{\be}{\begin{equation}}
\newcommand{\ee}{\end{equation}}
\newcommand{\bea}{\begin{eqnarray}}
\newcommand{\eea}{\end{eqnarray}}

\newcounter{num}

\begin{document}

\begin{titlepage}
%
%


%

\begin{centering}
\vspace{1cm}
{\Large {\bf Positivity Bounds on Higgs-Portal \\ \vspace{0.2cm} Freeze-in Dark Matter}} \\

\vspace{1.5cm}

{\bf Seong-Sik Kim$^{1,\ddagger}$, Hyun Min Lee$^{1,\dagger}$, and Kimiko Yamashita$^{2,\star}$ }
\vspace{.5cm}

{\it  $^1$Department of Physics, Chung-Ang University, Seoul 06974, Korea.}  \\[0.2cm]
{\it $^2$Department of Physics, Ibaraki University, Mito 310-8512, Japan.}

\vspace{.5cm}


\end{centering}
\vspace{2cm}

\begin{abstract}
\noindent
We consider the relic density and positivity bounds for freeze-in scalar dark matter with general Higgs-portal interactions up to dimension-8 operators. When dimension-4 and dimension-6 Higgs-portal interactions are proportional to mass squares for Higgs or scalar dark matter in certain microscopic models such as massive graviton, radion or general metric couplings with conformal and disformal modes, we can take the dimension-8 derivative Higgs-portal interactions to be dominant for determining the relic density via the 2-to-2 thermal scattering of the Higgs fields after reheating. We discuss the implications of positivity bounds for microscopic models. First, massive graviton or radion mediates attractive forces between Higgs and scalar dark matter and the resultant dimension-8 operators respect the positivity bounds. Second, the disformal couplings in the general metric allow for the subluminal propagation of graviton but violate the positivity bounds. We show that there is a wide parameter space for explaining the correct relic density from the freeze-in mechanism and the positivity bounds can curb out the dimension-8 derivative Higgs-portal interactions nontrivially in the presence of the similar dimension-8 self-interactions for Higgs and dark matter.  

\end{abstract}
\vspace{3cm}

\begin{flushleft} 
${}^{\ddagger}$Email: sskim.working@gmail.com\\
${}^{\dagger}$Email: hminlee@cau.ac.kr\\
${}^{\star}$Email: kimiko.yamashita.nd93@vc.ibaraki.ac.jp
\end{flushleft}

\end{titlepage}

\baselineskip 18pt

\newpage

\section{Introduction}\label{sec:introduction}

As there has been no convincing hint for physics beyond the Standard Model (SM) at the Large Hadron Collider (LHC) and other experiments, it has become important to make precision measurements of the Higgs properties and describe the deviations from the SM predictions in terms of higher dimensional operators in the effective field theories (EFTs). 

On the other hand, there is a lot of evidence for dark matter from orbital velocities of galaxies, gravitational lensing, Cosmic Microwave Background anisotropies, large scale structures, etc. However, the nature of dark matter has not been known. As dark matter is almost neutral under the gauge symmetries of the SM, the EFT description of the interactions between dark matter and the SM has drawn a lot of attention for direct and indirect detections and LHC searches, enabling us to treat the mediator interactions for dark matter in a model-independent way. Although we need to take into account the validity of the EFT description at some high energy scales, we can still decode the important information  for microscopic models for dark matter in the EFT approach.

The consistent conditions for EFTs are manifest through the properties of the S-matrix such as analyticity, unitarity, and Lorentz invariance. In particular, such consistency conditions give rise to positivity conditions in the forward limit of the scattering amplitudes at low energies \cite{Adams:2006sv,Pham:1985cr,Ananthanarayan:1994hf}. Recently, positivity bounds have been applied to the SMEFTs \cite{Zhang:2021eeo,Bi:2019phv,Remmen:2019cyz,Ghosh:2022qqq,Zhang:2020jyn,Yamashita:2020gtt,Li:2022aby,Li:2022rag} and the Higgs-portal derivative interactions for scalar dark matter \cite{Kim:2023pwf}. In the latter work, it has been shown that the positivity bounds can constrain the parameter space further for explaining the relic density for Weakly Interacting Massive Particles \cite{Kim:2023pwf}. As a by-product of looking for the origin of the dimension-8 derivative Higgs-portal couplings, it has been found that the correlations between dimension-4 and dimension-6 operators in a Ultra-Violet(UV) complete model such as massive graviton or radion are crucial for identifying the safe Higgs-portal models from direct detection bounds \cite{Kim:2023pwf}. 

In this article, we extend our analysis of the positivity bounds in Ref.~\cite{Kim:2023pwf} to the case for freeze-in scalar dark matter with general Higgs-portal interactions. We assume that a sufficiently large reheating temperature is achieved after a slow-roll inflation and scalar dark matter is never in thermal equilibrium. In this case, we search for the parameter space where the correct relic density for dark matter is explained through the freeze-in process with the 2-to-2 thermal scattering of the Higgs fields in the SM thermal plasma. Then, in the case with suppressed dimension-4 and dimension-6 Higgs-portal interactions, which are proportional to Higgs or scalar dark matter masses, we determine the relic density dominantly by the dimension-8 derivative Higgs-portal interactions and in turn show how the positivity bounds are complementary to constrain the parameter space further. We also discuss the implications of the positivity bounds for microscopic origins of the Higgs-portal interactions such as massive graviton, radion and conformal/disformal couplings.

The paper is organized as follows. 
We first present the general effective interactions in the Higgs-portal scenario for scalar dark matter, discuss the positivity bounds on them including the derivative self-interactions for Higgs and dark matter, and address the implications for the positivity bounds for massive graviton, radion and the extended metric tensor with conformal and disformal modes. Next we determine the relic density for dark matter from the thermal scattering of the Higgs fields after reheating and impose the positivity bounds in the parameter space where the correct relic density is obtained. There is one appendix for the details of the thermal averages of the production rates for scalar dark matter with the most general Higgs-portal interactions up to dimension-8 operators. Finally, conclusions are drawn.

\section{Positivity bounds on Higgs-portal couplings}\label{sec:eft}

We first consider the effective Lagrangian for a real scalar dark matter $\varphi$ and the Higgs doublet  $H$ in the Standard Model, up to dimension-8 operators, and review the positivity bounds on them~\cite{Kim:2023pwf}.
Then, we discuss the implications of positivity bounds for microscopic models.

\subsection{Positivity bounds}

The effective Higgs-portal dark matter Lagrangian is:

\bea
{\cal L}_{\rm Higgs-portal}= {\cal L}_1 +{\cal L}_2
\eea
with
\bea
{\cal L}_1 &=&-\frac{1}{6\Lambda^4} \Big(c_1 m_\varphi^4 \varphi^4 + 4c_2m^2_H |H|^4 + 8c'_2 m^2_H\lambda_H  |H|^6 + 4c^{\prime\prime}_2\lambda^2_H  |H|^8 \nonumber\\
&&\quad+ 4 c_3 m^2_H m^2_\varphi \varphi^2 |H|^2+4c'_3 \lambda_H  m^2_\varphi \varphi^2 |H|^4  \Big) \nonumber \\
&&+\frac{1}{6\Lambda^4} \bigg( d_1 m^2_\varphi \varphi^2 (\partial_\mu\varphi)^2 + 4d_2 m^2_H |H|^2 |D_\mu H|^2+ 4d'_2\lambda_H  |H|^4 |D_\mu H|^2  \nonumber \\
&&\quad+2d_3 m_\varphi^2 \varphi^2 |D_\mu H|^2+ 2d_4m^2_H |H|^2 (\partial_\mu\varphi)^2+ 2d'_4\lambda_H  |H|^4 (\partial_\mu\varphi)^2 \bigg),  \label{lag1} \\
{\cal L}_2&=& \frac{C^{(1)}_{H^2\varphi^2}}{\Lambda^4} O^{(1)}_{H^2\varphi^2}+ \frac{C^{(2)}_{H^2\varphi^2}}{\Lambda^4} O^{(2)}_{H^2\varphi^2} \nonumber \\
&&+\frac{ C_{\varphi^4}}{\Lambda^4} O_{\varphi^4}+\frac{C^{(1)}_{H^4}}{\Lambda^4} O^{(1)}_{H^4}+\frac{C^{(2)}_{H^4}}{\Lambda^4} O^{(2)}_{H^4} +\frac{C^{(3)}_{H^4}}{\Lambda^4} O^{(3)}_{H^4} \label{effdim8}
\eea
where the coefficients of all the dimension-4 and dimension-6 operators are dimensionless parameters, and  $C^{(1)}_{H^2\varphi^2}, C^{(2)}_{H^2\varphi^2}, C_{\varphi^4}$ and $C^{(1,2,3)}_{H^4}$  are the Wilson coefficients for the dimension-8 operators containing four derivatives listed in  Table \ref{tab:dim8_operators}, and $\Lambda$ is the cutoff scale. 

\begin{table}[hbt!]
\center
\begin{tabular}{| l  l |}
\hline
$O^{(1)}_{H^2\varphi^2} = (D_\mu H^\dagger D_\nu H)(\partial^\mu \varphi \partial^\nu \varphi)$
& $O^{(2)}_{H^2\varphi^2} = (D_\mu H^\dagger D^\mu H)(\partial_\nu \varphi \partial^\nu \varphi)$\\
\hline\hline
$O_{\varphi^4} = \partial_\mu \varphi \partial^\mu \varphi \partial_\nu \varphi \partial^\nu \varphi$ & \\
\hline\hline
$O^{(1)}_{H^4} = (D_\mu H^\dagger D_\nu H)(D^\nu H^\dagger D^\mu H)$
& $O^{(2)}_{H^4} = (D_\mu H^\dagger D_\nu H)(D^\mu H^\dagger D^\nu H)$ \\
$O^{(3)}_{H^4} = (D_\mu H^\dagger D^\mu H)(D_\nu H^\dagger D^\nu H)$&\\
\hline
\end{tabular}
\caption{Dimension-8 operators for Higgs and real scalar dark matter}
\label{tab:dim8_operators}
\end{table}

We assume a $Z_2$ symmetry for the scalar dark matter, so the effective Higgs-portal interactions include only even numbers of scalar dark matter particles.  
We can also add the extra higher dimensional terms such as $ \phi^8, \phi^6, \phi^4 |H|^2, \phi^4 |H|^4$, but they are irrelevant for positivity and dark matter phenomenology in the leading order perturbation theory. So, we drop them in the following discussion. 

We also remark that it is convenient to parametrize the dimension-4 and dimension-6 operators in terms of the suppression factors of dark matter and Higgs masses as in Eq.~(\ref{lag1}). The reason is the following.  Considering the UV complete models where more fundamental interactions of  mediator particles are proportional to mass squares for dark matter and Higgs masses, we can get the lower dimensional operators suppressed by powers of $m^2_\varphi/\Lambda^2$ and $m^2_H/\Lambda^2$, so the dimension-8 operators become dominant for relatively high energy processes such as freeze-in production of dark matter at a reheating temperature satisfying $m_\varphi, m_H\ll T_{\rm reh}\lesssim \Lambda$.

From the scattering amplitudes for superposed states of Higgs and dark matter scalars, we can impose the following positivity bounds 
on the dimension-8 derivative Higgs-portal couplings~\cite{Kim:2023pwf},
\begin{align}
&C^{(1)}_{H^4}  + C^{(2)}_{H^4} \geq 0,  \label{eq:bound1} \\ 
&C^{(1)}_{H^4}  + C^{(2)}_{H^4} + C^{(3)}_{H^4} \geq 0,\\ 
&C^{(2)}_{H^4} \geq 0, \\
&C^{(1)}_{H^2\varphi^2} \geq 0, \label{eq:portal1}\\
&C_{\varphi^4} \geq 0, \\
&4\sqrt{(C^{(1)}_{H^4}  + C^{(2)}_{H^4} + C^{(3)}_{H^4})C_{\varphi^4}}
\geq \left|C^{(1)}_{H^2\varphi^2} + 2C^{(2)}_{H^2\varphi^2}\right| - C^{(1)}_{H^2\varphi^2}. \label{eq:boundf}
\end{align}

We note that the positivity condition in Eq.~\eqref{eq:boundf} can be rewritten as $-C^{(1)}_{H^2\varphi^2}-2A\leq C^{(2)}_{H^2\varphi^2}\leq 2A$ with $A\equiv \sqrt{(C^{(1)}_{H^4}  + C^{(2)}_{H^4} + C^{(3)}_{H^4})C_{\varphi^4}}$.
First, in the case with $C^{(2)}_{H^2\varphi^2}=+1$ and $C^{(1)}_{H^2\varphi^2}\geq 0$, the positivity condition in Eq.~\eqref{eq:boundf} leads to $A\geq \frac{1}{2}$, which is the lower bound on the product of the dimension-8 derivative self-interactions for Higgs and scalar dark matter. But, if $C^{(2)}_{H^2\varphi^2}=-1$, the positivity condition in Eq.~\eqref{eq:boundf} gives rise to $C^{(1)}_{H^2\varphi^2}\geq 1-2A$. Then, small dimension-8 self-interactions for Higgs and scalar dark matter can be compatible with the positivity bounds. Either cases with $C^{(2)}_{H^2\varphi^2}=+1$ or $-1$ are phenomenologically viable, given that Higgs-portal interactions are feeble for freeze-in scenarios, but we focus on the case with $C^{(2)}_{H^2\varphi^2}<0$ in the later discussion.

\subsection{Microscopic models for dimension-8 operators}

The dimension-8 operators as well as the lower dimensional operators in Eqs.~(\ref{lag1}) and (\ref{effdim8}) can be originated from the exchanges of 
a massive spin-2 particle and/or a radion-like scalar particle~\cite{Lee:2013bua,Lee:2014caa,GDD}, so we review the relation of the positivity bounds for those models in Ref.~\cite{Kim:2023pwf}.
We also comment on the case where scalar dark matter has conformal and disformal couplings through the metric \cite{Brax:2020gqg}.

First, suppose that there is an exchange of the massive spin-2 particle between the SM Higgs and the scalar dark matter through the energy-momentum tensor, in the following form,
\bea
{\cal L}_{G} = -\frac{c_H}{M} \, G^{\mu\nu} T^H_{\mu\nu} - \frac{c_\varphi}{M}\, G^{\mu\nu} T^\varphi_{\mu\nu}
\eea
where $T^H_{\mu\nu},  T^\varphi_{\mu\nu}$ are the energy-momentum tensors for Higgs and dark matter, given by
\bea
T^H_{\mu\nu} &=&(D_\mu H)^\dagger D_\nu H + (D_\nu H)^\dagger D_\mu H   - g_{\mu\nu}  [g^{\rho\sigma} (D_\rho H)^\dagger D_\sigma H] \nonumber \\
&&+ g_{\mu\nu} (m_H^2 |H|^2+\lambda_H |H|^4), \\
T^\varphi_{\mu\nu} &=& \partial_\mu \varphi \partial_\nu\varphi  -\frac{1}{2} g_{\mu\nu} (g^{\rho\sigma} \partial_\rho \varphi\partial_\sigma \varphi) +\frac{1}{2} g_{\mu\nu} m_\varphi^2 \varphi^2.
\eea
Here, $c_H, c_\varphi$ are dimensionless couplings and $M$ is the suppression scale for the spin-2 interactions. 
Then, after integrating out the massive spin-2 particle, we obtain the effective Lagrangian for the SM Higgs and the scalar dark matter~\cite{Kim:2023pwf},
\bea
{\cal L}_{G,{\rm eff}}=\frac{1}{4m^2_GM^2} \bigg( 2 T_{\mu\nu}T^{\mu\nu}-\frac{2}{3} T^2 \bigg),
\eea
where $T_{\mu\nu}=c_H T^H_{\mu\nu}+c_\varphi T^\varphi_{\mu\nu}$, $T=c_H T^H+c_\varphi T^\varphi$ with $T^H=T^{H,\mu}_\mu$ and $T^\varphi=T^{\varphi,\mu}_\mu$.
Then, we get the correlations between the effective Higgs-portal couplings for dark matter at the matching scale, as follows,
\begin{align}
\frac{C^{(2)}_{H^2\varphi^2}}{\Lambda^4}&=-\frac{1}{3} \frac{C^{(1)}_{H^2\varphi^2}}{\Lambda^4}=-\frac{2c_H c_\varphi}{3m^2_G M^2}, \label{G1}\\
\frac{c'_3}{\Lambda^4}&=\frac{c_3}{\Lambda^4}=\frac{d_3}{\Lambda^4}=\frac{d_4}{\Lambda^4}=\frac{d'_4}{\Lambda^4}=\frac{1}{2}\frac{C^{(1)}_{H^2\varphi^2}}{\Lambda^4}.  \label{G2}
\end{align}

In this case, if the massive spin-2 particle induces an attractive force between the Higgs and scalar dark matter, namely, $c_H c_\varphi>0$, leading to $C^{(1)}_{H^2\varphi^2}>0$ \cite{Kim:2023pwf}, the positivity bounds for the Higgs-portal interactions in Eqs.~(\ref{eq:portal1}) and \eqref{eq:boundf} are ensured. We regard the model with massive spin-2 particle as being another effective theory with undetermined signs of the couplings to the energy-momentum tensors. However, as in the massive graviton or Kaluza-Klein graviton from the extra dimension \cite{Lee:2013bua,Lee:2014caa,GDD}, the massive spin-2 particle can give rise to a new attractive force, being compatible with the positivity bounds and providing a consistent UV complete models.

On the other hand,  the radion can be exchanged between the SM Higgs and the scalar dark matter through the trace of the energy-momentum tensor, as follows,  
\bea
{\cal L}_r= \frac{c^r_H}{\sqrt{6}M} r\, T^H +\frac{c^r_\varphi}{\sqrt{6}M} r\, T^\varphi
\eea
with radion couplings, $c^r_H, c^r_\varphi$.
Then, after integrating out the radion, we find the effective Lagrangian for  the SM Higgs and the scalar dark matter~\cite{Kim:2023pwf} as
\bea
{\cal L}_{r,{\rm eff}}=\frac{1}{12m^2_r M^2}\, T^2
\eea
with $T=c^r_H T^H+c^r_\varphi T^\varphi$.
which leads to the correlations between the effective Higgs-portal couplings, as follows,
\begin{align}
C^{(1)}_{H^2\varphi^2} &=0,\label{r1}\\
\frac{c'_3}{\Lambda^4}&=\frac{c_3}{\Lambda^4}=\frac{d_3}{\Lambda^4}=\frac{d_4}{\Lambda^4}=\frac{d'_4}{\Lambda^4}=-6\frac{C^{(2)}_{H^2\varphi^2}}{\Lambda^4}=-\frac{2c^r_H c^r_\varphi}{m^2_r M^2}. \label{r2}
\end{align}
In this case, if the radion induces an attractive force between the Higgs and scalar dark matter, namely, $c^r_H c^r_\varphi>0$, we get $C^{(2)}_{H^2\varphi^2}>0$ \cite{Kim:2023pwf}, but it is not necessary for positivity bounds in view of Eq.~(\ref{eq:boundf}).
As a result, we find that the effective self-couplings for Higgs and scalar dark matter up to dimension-8 operators can be obtained from the graviton/radion exchanges, and they are correlated by one parameter originating from a more fundamental theory.

Before closing this section, we also comment on the case with conformal and disformal couplings for scalar dark matter \cite{Brax:2020gqg}.
Suppose that the Riemannian geometry is generalized to the Finsler geometry with the couplings for scalar dark matter, 
\bea
{\tilde g}_{\mu\nu}=C g_{\mu\nu} +D \partial_\mu\varphi \partial_\nu\varphi
\eea 
where   $C$ and $D$ are called the conformal and disformal couplings, respectively, taking the following form at the leading order,
\bea
C &=& 1+ c^2\,\frac{\varphi^2}{M^2_{Pl}} + c_X\,\frac{\partial_\mu\varphi\partial^\mu\varphi}{M^4_{Pl}}, \\
D &=& \frac{d}{M^4}+ \frac{d}{M^4}\, {\tilde c}^2\,\frac{\varphi^2}{M^2_{Pl}}.
\eea
Here, $c, c_X, d, {\tilde c}$ are dimensionless parameters and  $M$ is the cutoff scale. Thus, the scalar dark matter couples to the Higgs through the modified metric, as follows, \cite{Brax:2020gqg}, 
\bea
{\cal L}_{\rm eff} &=& -\frac{1}{2}  ({\tilde g}_{\mu\nu}-g_{\mu\nu}) T^{\mu\nu}_H \nonumber \\
&=&-\frac{1}{2}(C-1) T^{H,\mu}_\mu -\frac{1}{2} D\,  \partial_\mu\varphi \partial_\nu\varphi \,T^{H,\mu\nu}
\eea
where $ T^{\mu\nu}_H $ is the energy-momentum tensor for the Higgs fields.
Then, ignoring the non-derivative interactions, i.e. $c={\tilde c}=0$, and taking $M=\Lambda$ and $c_X={\tilde c}_X M^4_{Pl}/\Lambda^4$, we get the effective interactions between the Higgs and scalar dark matter by
\bea
C^{(1)}_{H^2\varphi^2} =-d, \qquad C^{(2)}_{H^2\varphi^2}  = \frac{1}{2}d +{\tilde c}_X. 
\eea
Therefore, the positivity bound in Eq.~(\ref{eq:portal1}) implies that $d<0$, whereas ${\tilde c}_X$ can be either positive or negative. This result is in contrast to the causality condition, $d>0$, which comes from the sub-luminal propagation of the metric perturbations around the Finsler geometry \cite{Bekenstein:1992pj}.  In contrast, the positivity bounds with $d<0$ in our work stem from the perturbations of the SM Higgs which couple to the modified gravity sector, limiting the possibility of the Finsler geometry as a UV completion of the effective dimension-8 Higgs-portal interactions. 
Our result shows another example where the positivity bounds are violated, other than certain massive gravity models in the literature \cite{Bellazzini:2023nqj}.

\section{Dark matter relic density via Higgs-portal freeze-in}\label{sec:dm_relic}

In this section, we consider the constraints on the effective Higgs-portal couplings for the freeze-in production of dark matter. 
We discuss the interplay between the relic density of dark matter and the positivity bounds in constraining the dimension-8 derivative Higgs-portal couplings. For some related papers on the freeze-in production of dark matter through the exchanges of a massive graviton in the literature, we refer to Refs.~\cite{Bernal:2018qlk,Bernal:2020fvw,Brax:2020gqg}.

\subsection{Boltzmann equations for dark matter}

\begin{figure}[!t]
\begin{center}
 \includegraphics[width=0.25\textwidth,clip]{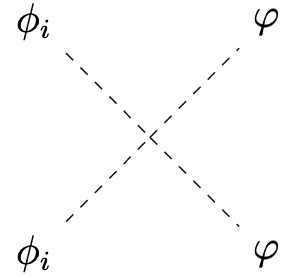}
\end{center}
\caption{Feynman diagrams for dark matter production due to effective Higgs-portal interactions. Here, $\phi_i(i=1,2,3,4)$ are the four real scalar fields of the Higgs doublet, and $\varphi$ is the real scalar dark matter. }
\label{fig:diagram_RelicD}
\end{figure}

We can determine the relic density by the freeze-in mechanism \cite{Dodelson:1993je,Hall:2009bx,Bernal:2017kxu} in our scenario. To this, we consider the production channels for scalar dark matter via $\phi_i \phi_i \to \varphi \varphi$ $(i=1,2,3,4)$, as shown in Fig.~\ref{fig:diagram_RelicD}. The effective Higgs-portal interactions relevant for the freeze-in mechanism are dimension-4 and dimension-6 Higgs-portal couplings, such as terms with $c_3$, $d_3$, and $d_4$ in Eq.~\eqref{lag1}, and the dimension-8 derivative Higgs-portal couplings also contribute. 

For the freeze-in mechanism to work, we need to require the Higgs-portal couplings to be small enough such that dark matter is decoupled from the thermal plasma. Thus, we impose the dimension-4 Higgs-portal operator with $c_3 m^2_H m^2_\varphi/\Lambda^4\lesssim 10^{-7}$ \cite{Choi:2019osi}, the dimension-6 Higgs-portal operators with $d_3 m^2_\varphi/\Lambda^4, d_4m^2_H/\Lambda^4\lesssim 1/(T^3_{\rm reh} M_{Pl})^{1/2}$, and the dimension-8 derivative Higgs-portal operators with $C^{(1,2)}_{H^2\varphi^2}/\Lambda^4\lesssim 1/(T^7_{\rm reh} M_{Pl})^{1/2}$, where $M_{Pl}= 2.4\times10^{18}$~GeV is the reduced Planck mass and $T_{\rm reh}$ is the reheating temperature. For $c_3, d_4, d_3, C^{(1,2)}_{H^2\varphi^2}={\cal O}(1)$, the minimum cutoff scale for the kinematic decoupling is determined by the dimension-8 operators, namely, $\Lambda\gtrsim (T^7_{\rm reh}M_{Pl})^{1/8}$. For instance, we need $\Lambda\gtrsim 10^{14.5}\,{\rm GeV}$ for $T_{\rm reh}=10^{14}\,{\rm GeV}$ and $\Lambda\gtrsim 10^{7.5}\,{\rm GeV}$ for $T_{\rm reh}=10^6\,{\rm GeV}$, so it is sufficient to take the cutoff scale to be larger than the reheating temperature by order of magnitude. Then, we can satisfy automatically the unitarity condition, which corresponds to $T_{\rm reh}\lesssim \Lambda$ in the limit of ignoring the masses for Higgs and dark matter.

 A comment on the running effects of the dimension-8 operators is in order. 
The positivity bounds can be set at the matching scale where new mediator particles in UV complete models are integrated out. However, we need to take into account the running effects of the dimension-8 operators when the 2-to-2 scattering processes such as freeze-in production take place at a much lower energy as compared to the matching scale to UV complete models. In this case, not only the dimension-8 self-interactions for Higgs \cite{Li:2022aby} but also the dimension-8 Higgs-portal interactions \cite{Kim:2023pwf} are subject to the logarithmic running effects, so small negative values of the low-energy Wilson coefficients of the dimension-8 Higgs-portal interactions are allowed by the positivity bounds \cite{Kim:2023pwf}. 
 
In the later discussion, we assume that the dimension-4 and dimension-6 operators are suppressed by the mass squares of Higgs or dark matter scalars, but the dimension-8 derivative Higgs-portal couplings are dominant for dark matter production at the high reheating temperature. In this case, the correlation between the freeze-in dark matter and the positivity bounds becomes more manifest. 

We take the representation of the SM Higgs doublet  $H$ in terms of four real scalar fields, $\phi_i (i=1,2,3,4)$,  as follows,
\begin{align}
H =\frac{1}{\sqrt{2}}\begin{pmatrix}
\phi_1 + i\phi_2 \\
\phi_3 + i\phi_4
\end{pmatrix}.
\end{align}
We assume that the electroweak symmetry is  unbroken during the freeze-in production of dark matter, so all the Higgs scalar fields contribute equally to the freeze-in processes. 

The number density for the dark matter, $n_\varphi$, is governed by the following Boltzmann equation,
\begin{align}
\dot{n}_{\varphi} + 3Hn_{\varphi} &= R(t)
\label{eq:boltzmann}
\end{align}
where $R(t)$ is the production rate for dark matter per unit volume and per unit time, the Hubble parameter during radiation domination is taken to $H = \sqrt{\frac{{\rho_R}}{3 M_{\mathrm{Pl}}}}=\sqrt{\frac{g_*\pi^2}{90}}\frac{T^2}{M_{\mathrm{Pl}}}$,
with  $\rho_R$ being the radiation energy density, and $g_*$ is the number of effective relativistic degrees of freedom.

Defining the dark matter yield as $Y_{\varphi} \equiv \frac{n_{\varphi}}{T^3}$,
 we can rewrite Eq.~\eqref{eq:boltzmann} as the Boltzmann equation for the dark matter yield $Y_{\varphi}$, in terms of the temperature $T$, as
\begin{align}
\frac{dY_{\varphi}}{dT}&= -\frac{R(T)}{H(T)T^4}.
\label{eq:boltzmann2}
\end{align}
Here, the production rate $R(T)$ for the process $A B \to C D$ is given by
\begin{align}
R(T) = g_\phi\int f_A f_B \frac{E_A E_B \mathrm{d}E_A \mathrm{d}E_B \mathrm{d}\cos{\theta_{AB}}}{1024\pi^6}\int|\mathcal{M}_{AB\to CD}|^2
\mathrm{d}\Omega_{AC}\sqrt{1-\frac{4m^2_\varphi}{s}}
\label{eq:rate}
\end{align}
where $g_\phi$ is the number of degrees of freedom for the Higgs fields, which is $g_\phi=4$,
\begin{align}
f_{A/B}  = \frac{1}{{\rm Exp}({E_{A/B}/T}) -1}
\label{eq:bose_d}
\end{align} 
are the occupancy distributions of the Higgs fields, ${\theta_{AB}}$ is the angle between the Higgs boson $A$ and $B$ in the laboratory frame,
$\mathrm{d}\Omega_{AC}$ is the solid angle between the Higgs boson $A$ and the scalar dark matter $C$ in the center of mass (COM) frame, and
$|\mathcal{M}_{AB\to CD}|^2$ is the squared matrix element of the process $AB\to CD$.
Here, we took  each initial Higgs field to be massless in the unbroken phase. Here, we took  each initial Higgs field to be massless in the unbroken phase. When the dark matter mass is larger than the temperature, namely, $E_A, E_B\gtrsim m_\varphi>T$, leading to the Boltzmann suppression, $f_{A,B}\simeq e^{-E_{A,B}/T}$.

\subsection{Dark matter relic density and positivity bounds}

We continue to present the main results of our work on freeze-in production in detail.
Moreover, we add the positivity bounds on the dimension-8 derivative Higgs-portal couplings. 

First, in terms of the coefficients of the Higgs-portal effective operators, we obtain the squared scattering amplitude for $\phi_i \phi_i \to \varphi \varphi (i = 1, 2, 3, 4)$, $|M_{\phi_i\phi_i\to\varphi\varphi}|^2 $,  as follows,
\begin{align}
|\mathcal{M}_{\phi_i\phi_i\to\varphi\varphi}|^2 &= \frac{1}{576\Lambda^8}
\bigg(
3 (C^{(1)}_{H^2\varphi^2} + 2  C^{(2)}_{H^2\varphi^2}) s^2
  + 6  C^{(1)}_{H^2\varphi^2} m^4_\varphi 
  - 8 (2c_3 - d_4) m^2_H m^2_\varphi  \nonumber\\
& + 6  C^{(1)}_{H^2\varphi^2} (t^2 - 2 m^2_\varphi t) 
  + 2 s \left[3 C^{(1)}_{H^2\varphi^2} t  - (3 C^{(1)}_{H^2\varphi^2} + 6  C^{(2)}_{H^2\varphi^2} + 2 d_3) m^2_\varphi 
  -2 d_4 m^2_H \right] \bigg)^2
\label{eq:amp2}
\end{align}
where the symmetric factors for identical particles in initial states (i.e., $\phi_i\phi_i$ pairs) and final states (i.e., dark matter particle pair, $\varphi\varphi$) are included.
The Mandelstam variables, $s$ and $t$, are related to the angle between initial states in the lab frame, $\theta_{AB}$, and  the scattering angle $\theta_{AC}$ in the COM frame, as follows, 
\begin{align}
s &= 2E_A E_B (1- \cos{\theta_{AB}}), \label{ss} \\
t &= \frac{s}{2}\left(\sqrt{1-\frac{4m^2_\varphi}{s}}\cos{\theta_{AC}} -1 \right) + m^2_\varphi.
\end{align}
We can take $\mathrm{d}\Omega_{AC} = 2\pi \mathrm{d}\cos{\theta_{AC}}$ due to the azimuthal symmetry in the COM frame.
Here, we note that $s$ appearing in $t$ is taken in the COM frame, but its value is identical to the one in the lab frame, so we take the form of $s$ in the lab frame in Eq.~(\ref{ss}), which is appropriate for the thermal averages.

We comment on the scattering angle dependence of the scattering amplitude in connection to the partial UV completion of the dimension-8 operators.
Taking $s, t\gg m^2_\varphi, m^2_H$, we can approximate the  squared scattering amplitude in Eq.~(\ref{eq:amp2}) to
\bea
|\mathcal{M}_{\phi_i\phi_i\to\varphi\varphi}|^2\simeq \frac{1}{576\Lambda^8} \bigg[3 (C^{(1)}_{H^2\varphi^2} + 2  C^{(2)}_{H^2\varphi^2}) s^2+6 C^{(1)}_{H^2\varphi^2} \, t(t+s) \bigg]^2.
\eea
We note that the terms containing $c_3, d_3, d_4={\cal O}(1)$ in  Eq.~(\ref{eq:amp2}) are neglected in the massless limit for Higgs and dark matter.
Then, in the case with the exchanges of the massive graviton for dark matter, we can set $C^{(1)}_{H^2\varphi^2}=-3C^{(2)}_{H^2\varphi^2}$, for which the squared scattering amplitude in Eq.~(\ref{eq:amp2}) is proportional to $|\mathcal{M}_{\phi_i\phi_i\to\varphi\varphi}|^2\propto s^4 (1-3\cos^2\theta)^2$, being $(d\text{-wave})^2$, in the massless limit for Higgs and dark matter. On the other hand, in the case with the exchanges of the radion for dark matter, we can take $C^{(1)}_{H^2\varphi^2}=0$, for which the squared scattering amplitude in Eq.~(\ref{eq:amp2}) is independent of the scattering angle, being 
 $(s\text{-wave})^2$.

\begin{figure}[!t]
\begin{center}
 \includegraphics[width=0.40\textwidth,clip]{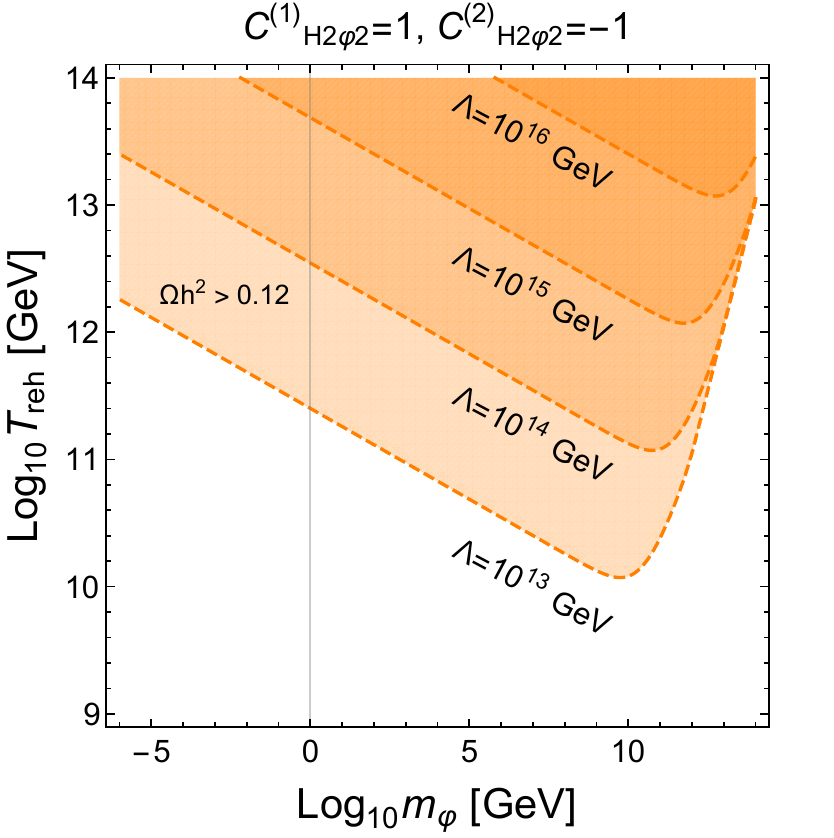}\,\,
\includegraphics[width=0.40\textwidth,clip]{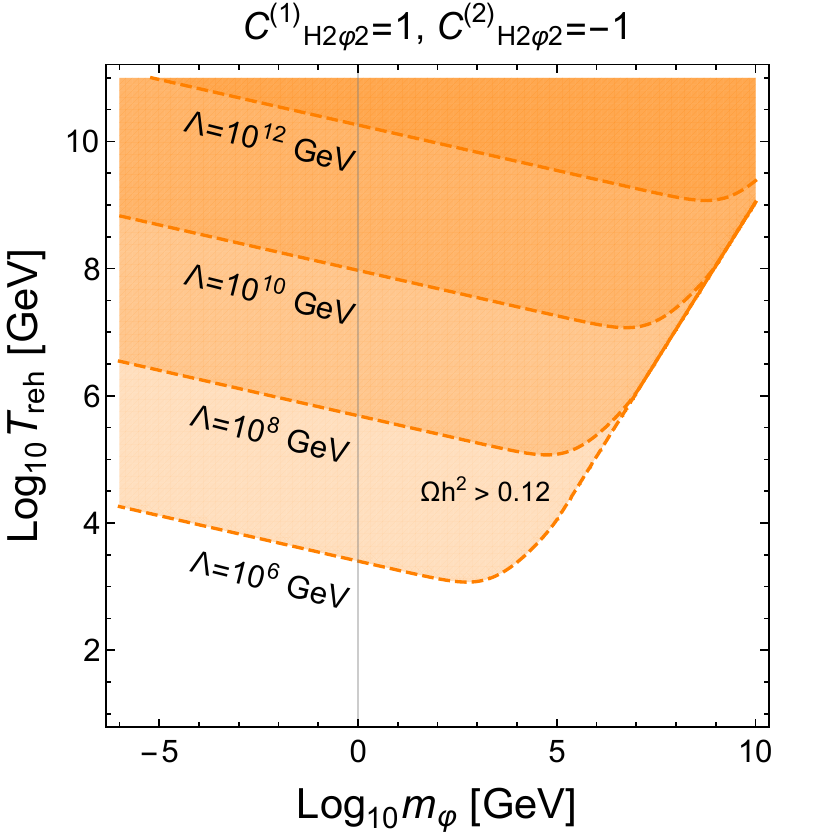}
\end{center}
\caption{Parameter space for $m_{\varphi}$ vs  $T_\mathrm{reh}$, satisfying the positivity bounds and the relic density  for dark matter.
We took the values of $\Lambda = 10^{13, 14, 15, 16}\, {\rm GeV}(10^{6, 8, 10, 12} \,{\rm GeV}) $ for the left (right) plots and
$C^{(1)}_{H^2\varphi^2} = 1$ and $C^{(2)}_{H^2\varphi^2} = -1$ in common.
We took the dark matter mass only above 1 keV because of the Lyman$-\alpha$ bounds on warm dark matter~\cite{Decant:2021mhj}.
}
\label{fig:RelicD2}
\end{figure}

As discussed previously, dimension-4 and dimension-6 terms are suppressed by the mass squares for Higgs and dark matter, so the leading contribution to the freeze-in production is through dimension-8 operators, being dominated near the reheating temperature. Moreover, we bound the reheating temperature to be smaller than the cutoff scale in the effective theory such that even higher dimensional operators make minor contributions to the freeze-in production in the UV.  More explicitly, we consider the dimension$-(4+2n)$ Higgs-portal interactions with $2n$ derivatives in the effective theory below the cutoff scale $\Lambda$. Then, those higher dimensional operators give rise to the leading scattering rate for dark matter production scaling by $R(T)\sim \frac{T^{12}}{\Lambda^8}\big(\frac{T}{\Lambda}\big)^{4(n-2)}$, so the resultant relic abundance for dark matter is determined by the reheating temperature to be $Y_\varphi\sim \frac{1}{4n-1}\,\frac{T^7_{\rm reh}M_{Pl}}{\Lambda^8}\big(\frac{T_{\rm reh}}{\Lambda}\big)^{4(n-2)}$ by integrating Eq.~(\ref{eq:boltzmann2}) between $T=T_{\rm reh}$ and $T=m_H$. As a result, the higher dimensional operators with more than four derivatives, namely, $n>2$, are sub-dominant for the freeze-in production as compared to the dimension-8 operators, as far as the reheating temperature is below the cutoff scale, namely, $T_{\rm reh}\lesssim \Lambda$.

As a consequence, from the results in the Appendix on the production rate for dark matter and the relic abundance, for $T_{\rm reh}\gg m_\varphi, m_H$, we find  that dark matter is produced mostly around the reheating temperature.
Thus, we get the approximate formula for the relic abundance in our scenario as
\begin{align}
\Omega_\varphi h^2 
&\simeq 0.12  \left(\frac{m_\varphi}{1~\mathrm{TeV}}\right)\left(\frac{T_\mathrm{reh}}{10^{11}~\mathrm{GeV}}\right)^7
\left(\frac{10^{13}~\mathrm{GeV}}{\Lambda}\right)^8 \nonumber \\
&\quad\times \left(\frac{14}{5} (C^{(1)}_{H^2\varphi^2})^2+16 C^{(1)}_{H^2\varphi^2} C^{(2)}_{H^2\varphi^2}+24 (C^{(2)}_{H^2\varphi^2})^2\right).
\end{align}

In Fig.~\ref{fig:RelicD2}, we depict the parameter space in $m_{\varphi}$ vs  $T_\mathrm{reh}$, satisfying the relic density for dark matter. We chose  values of the suppression scale for the Higgs-portal effective interactions to $\Lambda = 10^{13, 14, 15, 16} \,{\rm GeV}$ on the left plot, $\Lambda=10^{6, 8, 10, 12} \,{\rm GeV}$ on the right plot, but we took $C^{(1)}_{H^2\varphi^2} = 1$ and $C^{(2)}_{H^2\varphi^2} = -1$ for both plots.
We have also taken into account the Boltzmann suppression for dark matter production when $m_{\varphi}\gtrsim T_{\mathrm{reh}}$.
We also remark that the dark matter masses below about 1 keV is disfavored by Lyman$-\alpha$ constraints~\cite{Decant:2021mhj}.
The larger cutoff scale $\Lambda$ or the smaller reheating temperature, the heavier dark matter mass is preferred.
Moreover, the lower cutoff scale $\Lambda$, the smaller reheating temperature $T_{\mathrm{reh}}$ is needed for the correct relic density.
Regarding the reheating temperature as being a free parameter, we find that the wide ranges of dark matter masses and the cutoff scales are consistent with the dark matter relic density at present.

As discussed in the end of Section 2.1, we recall that $C^{(2)}_{H^2\varphi^2} <0$ is not a necessary condition, because the positivity bound in Eq.~(\ref{eq:boundf}) involving $C^{(2)}_{H^2\varphi^2}$ depends on $C^{(1)}_{H^2\varphi^2}$ and the other dimension-8 operators such as $C^{(i)}_{H^4}(i=1,2,3)$ and $C_{\varphi^4}$. For instance, in the case of the radion couplings, we have $C^{(1)}_{H^2\varphi^2}=0$, so there is only an upper bound on the absolute value of $C^{(2)}_{H^2\varphi^2}$ when $(C^{(1)}_{H^4}+C^{(2)}_{H^4}+C^{(3)}_{H^4}) C_{\varphi^4}\neq 0$, but the sign of  $C^{(2)}_{H^2\varphi^2}$ is unconstrained. However, the case with a sizable positive value of $C^{(2)}_{H^2\varphi^2}$ is allowed only in the limited parameter space with non-vanishing self-interactions for Higgs and dark matter, as will be shown shortly. 
	
\begin{figure}[!t]
\begin{center}
\includegraphics[width=0.40\textwidth,clip]{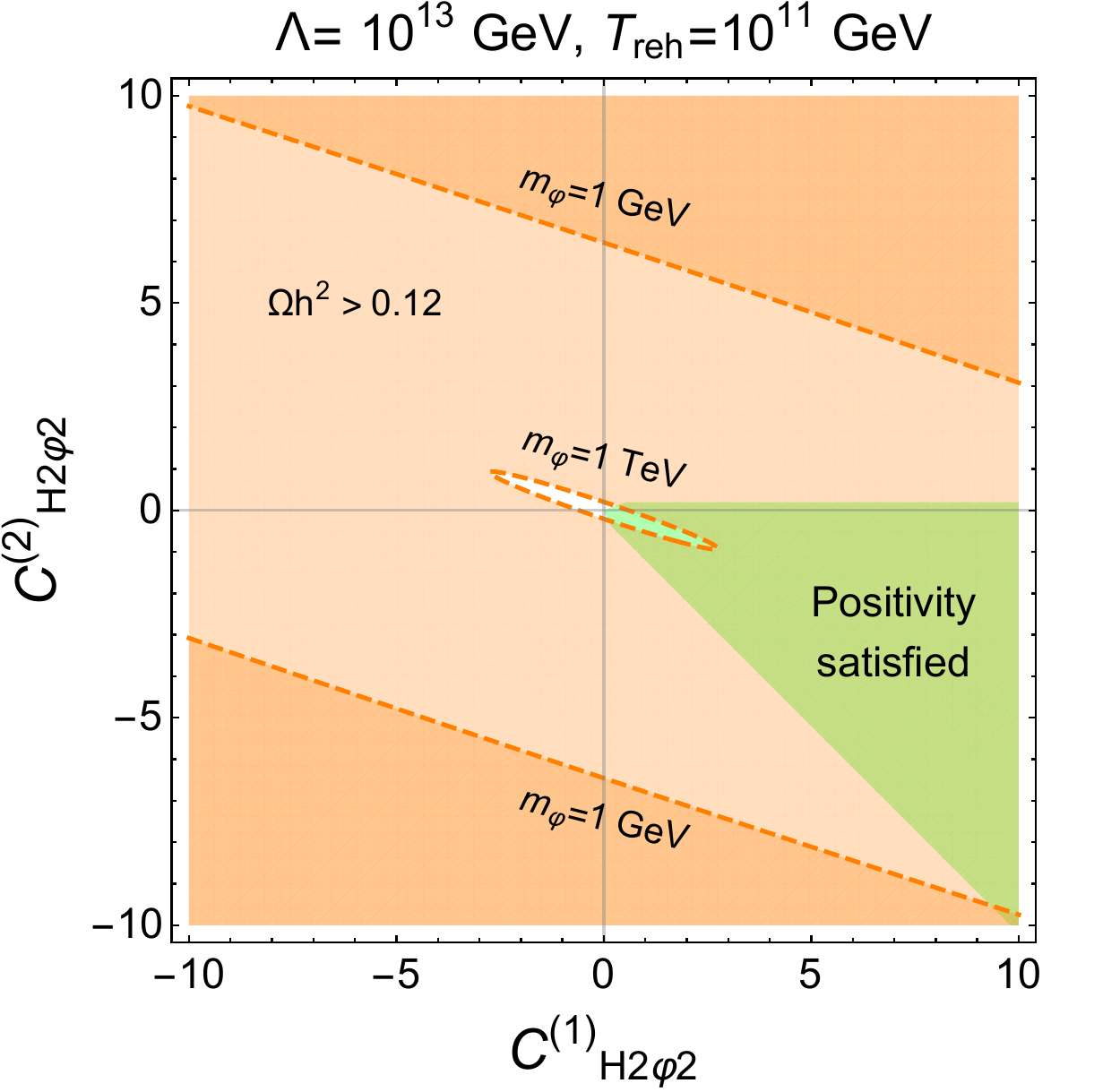} \,\,
\includegraphics[width=0.40\textwidth,clip]{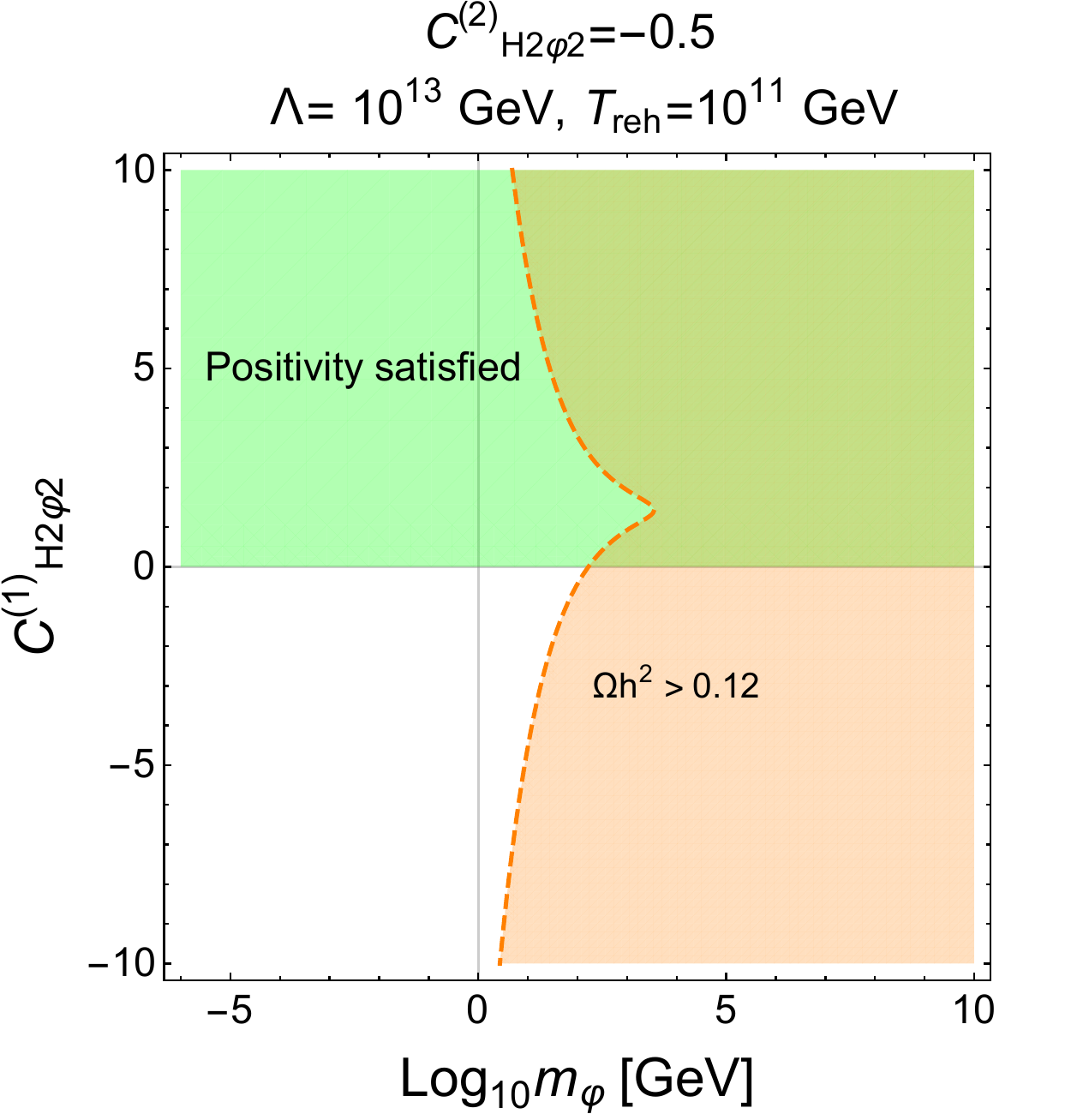}
\end{center}
\caption{Left: Parameter space for $C^{(1)}_{H^2\varphi^2}$ vs $C^{(2)}_{H^2\varphi^2}$, satisfying positivity and relic density.
We chose the dark matter mass to $m_{\varphi}=1\,{\rm GeV}$ and $1\,{\rm TeV}$.
For $A\equiv \sqrt{(C^{(1)}_{H^4}  + C^{(2)}_{H^4} + C^{(3)}_{H^4})C_{\varphi^4}}=0.1$,
the positivity bounds in Eqs.~\eqref{eq:portal1} and \eqref{eq:boundf} are satisfied in green.
Right: Parameter space for $m_{\varphi}$ vs $C^{(1)}_{H^2\varphi^2}$, satisfying positivity and relic density, 
for $C^{(2)}_{H^2\varphi^2}=-0.5$.
The positivity bounds in Eq.~\eqref{eq:portal1} are satisfied in green regions.
In common for all the plots: we took $\Lambda = 10^{13}$~GeV and $T_\mathrm{reh} = 10^{11}$~GeV, and dark matter relic density is overproduced in orange regions, namely, $\Omega h^2>0.12$, and it saturates the observed value along the boundary of the orange region.}
\label{fig:RelicD}
\end{figure}

\begin{figure}[!t]
\begin{center}
\includegraphics[width=0.40\textwidth,clip]{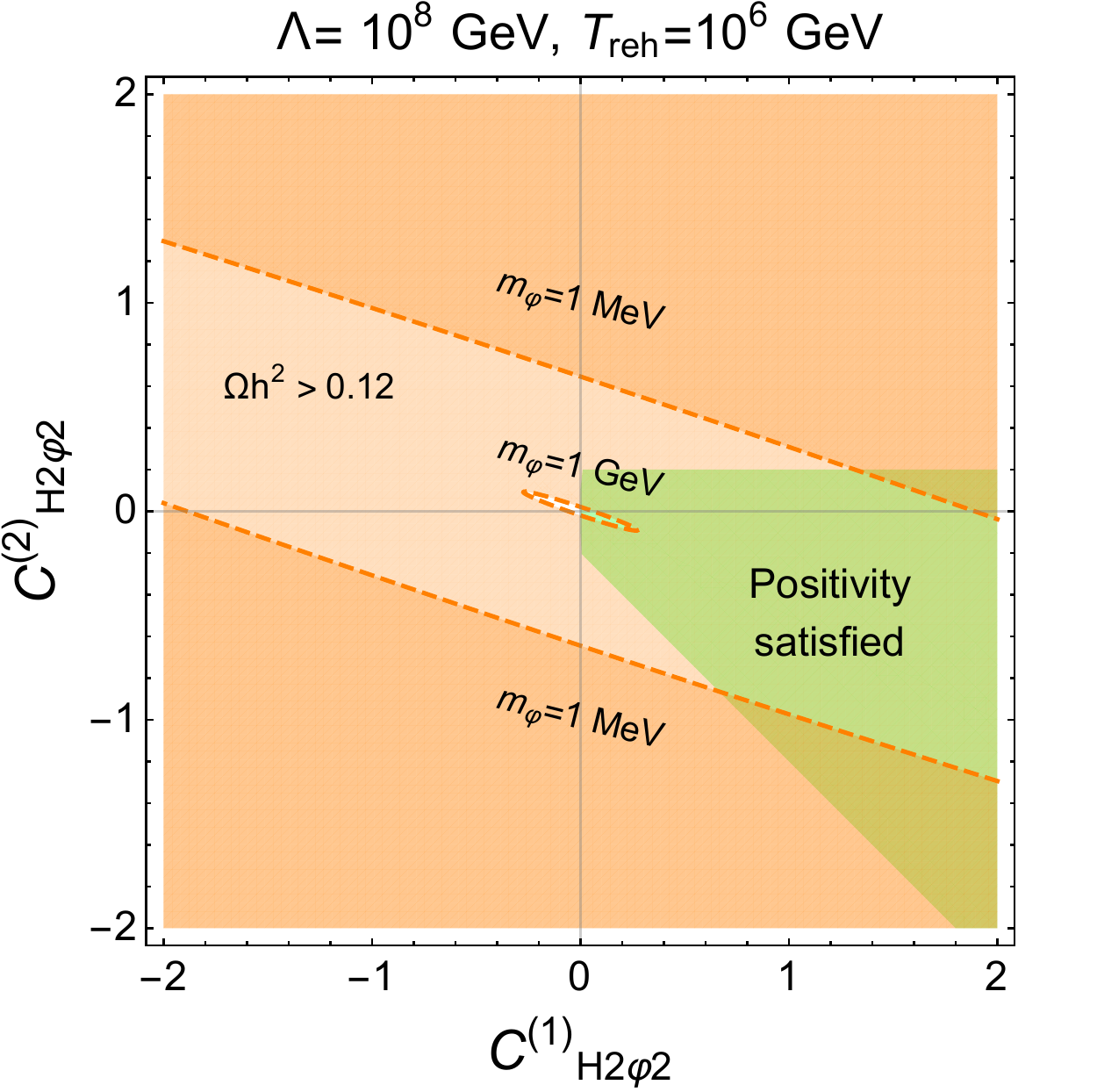} \,\,
\includegraphics[width=0.40\textwidth,clip]{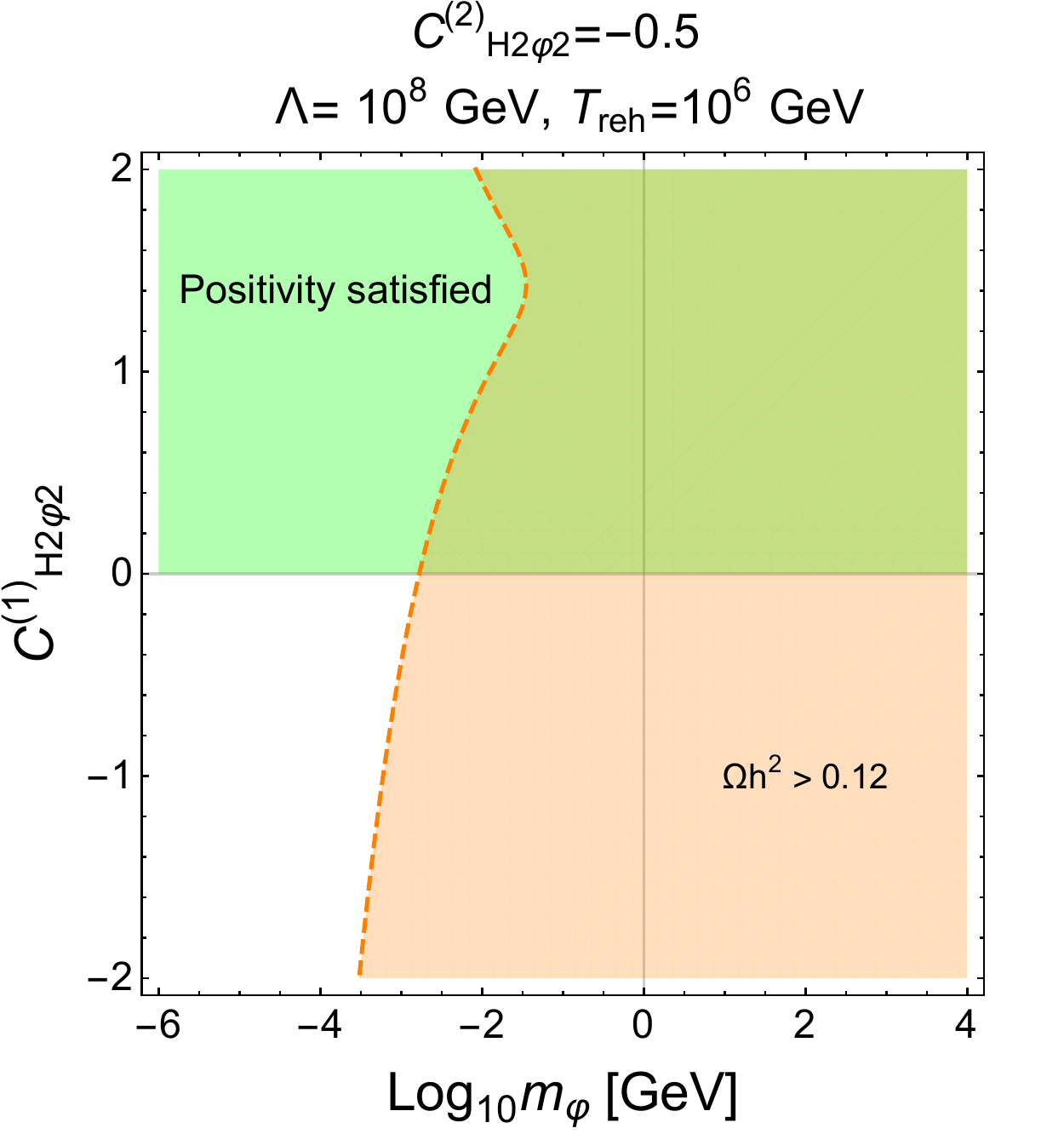}
\end{center}
\caption{The same as in Fig.~\ref{fig:RelicD}, except that $\Lambda = 10^{8}$~GeV and $T_\mathrm{reh} = 10^{6}$~GeV.}
\label{fig:RelicD3}
\end{figure}

In Fig.~\ref{fig:RelicD}, we show the parameter space for $C^{(1)}_{H^2\varphi^2}$ and $C^{(2)}_{H^2\varphi^2}$ in the left plot and for $m_{\varphi}$ and $C^{(1)}_{H^2\varphi^2}$ in the right plot, satisfying the relic density for dark matter and positivity bounds. 
We fixed $\Lambda = 10^{13}~$GeV and $T_\mathrm{reh} = 10^{11}~$GeV for both figures.  The relic density with $\Omega h^2<0.12$ is achieved outside the orange regions, whereas the observed relic density, $\Omega h^2=0.12$, is explained for $m_\varphi=1-10^3\,{\rm GeV}$ along the boundary of the orange region. The region satisfying the positivity bounds are shown in green. 
For the left plot of  Fig.~\ref{fig:RelicD}, we also took the combination of the dimension-8 derivative self-interactions for Higgs and dark matter by $A\equiv \sqrt{(C^{(1)}_{H^4}  + C^{(2)}_{H^4} + C^{(3)}_{H^4})C_{\varphi^4}}=0.1$. We can also allow for a larger value of $A$ as far as the cutoff $\Lambda$ is sufficiently large to avoid the experimental constraints at the LHC. Hence, the positivity bounds in Eq.~\eqref{eq:boundf} are satisfied even for $C^{(2)}_{H^2\varphi^2}>0$ in the left plot of Fig.~\ref{fig:RelicD}, although dark matter is overproduced for $m_\varphi\simeq 1$~TeV.
For $A=0.1$, the positivity bounds exclude a large portion of the parameter space with small dark matter masses below $1\,{\rm TeV}$ for $|C^{(1)}_{H^2\varphi^2}| < 10$. Moreover, in Fig.~\ref{fig:RelicD3}, we also show the similar results for $\Lambda = 10^{8}~$GeV and $T_\mathrm{reh} = 10^{6}~$GeV, but the correct relic density is now achieved for $m_\varphi=10^{-3}-1\,{\rm GeV}$.

As shown in the right plot of Figs.~\ref{fig:RelicD} or \ref{fig:RelicD3}, there are destructive interferences between $C^{(1)}_{H^2\varphi^2}$ and $C^{(2)}_{H^2\varphi^2}$  when they take the opposite signs, favoring a relatively larger dark matter mass around $C^{(1)}_{H^2\varphi^2} \gtrsim -2C^{(2)}_{H^2\varphi^2}$.
We checked that a complete cancellation between the different dimension-8 operators does not occur.

\section{Conclusions}\label{sec:summary}

We considered the interplay of the relic density and positivity bounds for constraining the general Higgs-portal interactions for scalar dark matter up to dimension-8 operators. We focused on the production of dark matter by the freeze-in mechanism where dark matter is produced by the 2-to-2 thermal scattering of the Higgs fields after reheating. As a result, we showed that when dimension-4 and dimension-6 operators are suppressed by mass squares for Higgs or dark matter, the dimension-8 derivative Higgs-portal interactions determine the relic density dominantly. 
We found that when massive graviton or radion is exchanged between Higgs and scalar dark matter as attractive forces, the resultant effective dimension-8 operators are consistent with the positivity bounds. However, the disformal couplings in the general metric can respect the subluminal propagation of graviton but violate the positivity bounds.

Taking into account both the freeze-in condition and the unitarity condition during the production of dark matter, we found that there is a lot of the parameter space satisfying the relic density for a sufficiently large reheating temperature. For instance, we obtained the correct relic density with dark matter masses, $m_\varphi=1-10^3\,{\rm GeV}$ and $m_\varphi=10^{-3}-1\,{\rm GeV}$, for $(\Lambda,T_\mathrm{reh})= (10^{13},10^{11})\,{\rm GeV}$ and $(\Lambda,T_\mathrm{reh})= (10^{8},10^{6})\,{\rm GeV}$, respectively.  We also illustrated how the positivity bounds can constrain the parameter space for the dimension-8 derivative Higgs-portal interactions, otherwise unconstrained by the relic density only.  Therefore, the positivity bounds are useful for identifying the consistent effective Higgs-portal interactions at low energies, for instance, coming from the couplings of massive graviton, radion as well as the extended metric interactions with conformal and disformal modes.

\def\theequation{A.\arabic{equation}}

\setcounter{equation}{0}

\vskip0.8cm
\noindent
{\Large \bf Appendix: Dark matter production rates}
\vskip0.3cm

We introduce the dark matter production rate for $\phi\phi\to \varphi\varphi$ considered in the text. (Here, we suppress the Higgs index $i$ for $\phi$.)
Then, we also compute the dark matter relic abundance in our scenario.

In general, we take the squared scattering amplitude for $\phi\phi\to \varphi\varphi$ in terms of the Mandelstam variables, $s$ and $t$, in the momentum expansion, as follows,
\begin{align}
|\mathcal{M}|^2 = 
\sum_{0 \le i+j \le 4} c_{i, j} \frac{s^i t^j}{M^{2(i+j)}}.\label{eq:sche_amp2}
\end{align}
Here, we introduced the suppression mass scale $M$ just for making the 2-to-2 scattering amplitude dimensionless in four dimensions.
Then, performing the thermal averages for the initial states and the phase space integrals for the final states in Eq.~\eqref{eq:rate}, we obtain the production rate for each term of Eq.~\eqref{eq:sche_amp2}, as follows,
\begin{align}
|\mathcal{M}|^2 &\to R(T); \nonumber\\
c_{0,0} &\to c_{0,0}\frac{T^4}{4608 \pi},\\
c_{1,0} \frac{s}{M^2} &\to \frac{c_{1,0}}{M^2} \frac{T^6 \zeta (3)^2}{16 \pi ^5},\\
c_{2,0} \frac{s^2}{M^4} &\to \frac{c_{2,0}}{M^4} \frac{\pi ^3 T^8}{5400},\\
c_{3,0} \frac{s^3}{M^6} &\to \frac{c_{3,0}}{M^6} \frac{72 T^{10} \zeta (5)^2}{\pi ^5},\\
c_{4,0} \frac{s^4}{M^8} &\to \frac{c_{4,0}}{M^8} \frac{128 \pi ^7 T^{12}}{19845},\\
c_{0,1} \frac{t}{M^2} &\to \frac{c_{0,1}}{M^2} \frac{\pi ^4 m_{\varphi}^2 T^4-144 T^6 \zeta (3)^2}{4608 \pi ^5},\\
c_{0,2} \frac{t^2}{M^4} &\to \frac{c_{0,2}}{M^4} 
\left(\frac{m_{\varphi}^4 T^4}{4608 \pi }-\frac{m_{\varphi}^2 T^6 \zeta (3)^2}{12 \pi ^5}+\frac{\pi ^3 T^8}{16200}\right),\\
c_{0,3} \frac{t^3}{M^6} &\to \frac{c_{0,3}}{M^6} 
\left(\frac{m_{\varphi}^6 T^4}{4608 \pi }-\frac{5 m_{\varphi}^4 T^6 \zeta (3)^2}{32 \pi ^5}+\frac{\pi ^3 m_{\varphi}^2 T^8}{3600}-\frac{18 T^{10} \zeta (5)^2}{\pi ^5}\right),\\
c_{0,4} \frac{t^4}{M^8} &\to \frac{c_{0,4}}{M^8}
\left(\frac{m_{\varphi}^8 T^4}{4608 \pi }-\frac{m_{\varphi}^6 T^6 \zeta (3)^2}{4 \pi ^5}+\frac{7 \pi ^3 m_{\varphi}^4 T^8}{9000}-\frac{576 m_{\varphi}^2 T^{10} \zeta (5)^2}{5 \pi ^5}+\frac{128 \pi ^7 T^{12}}{99225}\right),\\
c_{1,1}\frac{st}{M^4}&\to \frac{c_{1,1}}{M^4}
\left(\frac{m_{\varphi}^2 T^6 \zeta (3)^2}{16 \pi ^5}-\frac{\pi ^3 T^8}{10800}\right),\\
c_{2,1}\frac{s^2t}{M^6}&\to \frac{c_{2,1}}{M^6}
\left(\frac{\pi ^3 m_{\varphi}^2 T^8}{5400}-\frac{36 T^{10} \zeta (5)^2}{\pi ^5}\right),\\
c_{3,1}\frac{s^3t}{M^8}&\to \frac{c_{3,1}}{M^8}
\left(\frac{72  m_{\varphi}^2 T^{10}\zeta (5)^2}{\pi ^5}- \frac{64\pi^7T^{12}}{19845}\right),\\
c_{1,2}\frac{st^2}{M^6}&\to \frac{c_{1,2}}{M^6}
\left(\frac{m_{\varphi}^4 T^6 \zeta (3)^2}{16 \pi ^5}-\frac{\pi ^3 m_{\varphi}^2 T^8}{4050}+\frac{24 T^{10} \zeta (5)^2}{\pi ^5}\right),\\
c_{2,2}\frac{s^2t^2}{M^8}&\to \frac{c_{2,2}}{M^8}
\left(\frac{\pi ^3 m_{\varphi}^4 T^8}{5400}-\frac{96 m_{\varphi}^2 T^{10} \zeta (5)^2}{\pi ^5}+\frac{128 \pi ^7 T^{12}}{59535}\right),\\
c_{1,3}\frac{st^3}{M^8}&\to \frac{c_{1,3}}{M^8}
\left(-\frac{m_{\varphi}^6 T^6 \zeta (3)^2}{16 \pi ^5}-\frac{\pi ^3 m_{\varphi}^4 T^8}{2160}+\frac{108 m_{\varphi}^2 T^{10} \zeta (5)^2}{\pi ^5}-\frac{32 \pi ^7 T^{12}}{19845}\right).
\end{align}

We remark that the production rate for Eq.~\eqref{eq:sche_amp2} becomes, in the relativistic limit for dark matter,
\begin{align}
R(T) = \sum_{0\leq i+j} c_{i,j} (-1)^j \frac{2^{2(i+j)}\Gamma^2(i+j+2)\zeta^2(i+j+2)T^{2(i+j+2)}}{128\pi^5(j+1)(i+j+1)M^{2(i+j)}},
\end{align}
which is in agreement with \cite{Brax:2020gqg}.  

After integrating the Boltzmann equation in Eq.~\eqref{eq:boltzmann2} between the reheating temperature $T_{\rm reh}$ and the temperature $T=m_H$ when the Higgs fields are decoupled,  we get  the dark matter yield as
\begin{align}
Y_\varphi(m_H) &= Y_\varphi(T_\mathrm{reh}) + \frac{g_\phi}{\sqrt{g_*}}
\bigg[F_7(T^7_\mathrm{reh} - m_H^7)
+ F_5(T^5_\mathrm{reh} - m_H^5)
+ F_3(T^3_\mathrm{reh} - m_H^3)  \nonumber \\
& \, \, \, \, + F_1(T_\mathrm{reh} - m_H)
+ F_{-1}\left(\frac{1}{T_\mathrm{reh}} - \frac{1}{m_H}\right)
\bigg]
\label{eq:yield}
\end{align}
where
\begin{align}
F_7 &= \frac{32 \sqrt{\frac{2}{5}} \pi ^6 M_{Pl}(12 c_{0,4}+5 (-3 c_{1,3}+4 c_{2,2}-6 c_{3,1}+12 c_{4,0}))}{138915M^8},\\
F_5 &= -\frac{18 \sqrt{\frac{2}{5}} M_{Pl}\zeta (5)^2 \left(5 M^2 (3 c_{0,3}-4 c_{1,2}+6 (c_{2,1}-2 c_{3,0}))\right)}{5 \pi ^6 M^8}\nonumber\\
&\, \, -\frac{18 \sqrt{\frac{2}{5}} M_{Pl}\zeta (5)^2 \left(2 m_{\varphi}^2 (48 c_{0,4}-5 (9 c_{1,3}-8 c_{2,2}+6 c_{3,1}))\right)}{5 \pi ^6 M^8},\\
F_3 &= \frac{\pi ^2 M_{Pl}\left(5 M^4 (2 c_{0,2}-3 c_{1,1}+6 c_{2,0})+5 M^2 m_{\varphi}^2 (9 c_{0,3}-8 c_{1,2}+6 c_{2,1})\right)}{16200 \sqrt{10}M^8}\nonumber\\
&\, \, \, \,  +\frac{\pi ^2 M_{Pl}\left(3 m_{\varphi}^4 (42 c_{0,4}-25 c_{1,3}+10 c_{2,2})\right)}{16200 \sqrt{10}M^8}, \\
F_1 &= \frac{\sqrt{\frac{5}{2}} M_{Pl}\zeta (3)^2 \left(-3 M^6 (c_{0,1}-2 c_{1,0})+2 M^4 m_{\varphi}^2 (3 c_{1,1}-4 c_{0,2})+3 M^2 m_{\varphi}^4 (2 c_{1,2}-5 c_{0,3}))\right)}{16 \pi ^6M^8}\nonumber\\
&\, \, \, \, + \frac{\sqrt{\frac{5}{2}} M_{Pl}\zeta (3)^2 \left(6 m_{\varphi}^6 (c_{1,3}-4 c_{0,4})\right)}{16 \pi ^6M^8}, \\
F_{-1} &= -\frac{\sqrt{\frac{5}{2}} M_{Pl}\left(c_{0,0} M^8+c_{0,1} M^6 m_{\varphi}^2+c_{0,2} M^4 m_{\varphi}^4+c_{0,3} M^2 m_{\varphi}^6+c_{0,4} m_{\varphi}^8\right)}{768 \pi ^2M^8}.
\end{align}
Here, $Y_\varphi(T_\mathrm{reh})$ is the dark matter yield at the reheating temperature.
We note that we can fix $g_*$ to the value at the temperature which dominates the integration of the Boltzmann equation, namely, $g_*=g_*(T_{\rm reh})$ for the UV freeze-in case with $T_{\rm reh}\gg m_H$.

From the amplitude squared for the scattering $\phi\phi\to \varphi\varphi$, given in Eq.~\eqref{eq:amp2}, we obtain
\begin{align}
F_7 &= \frac{2 \sqrt{\frac{2}{5}} \pi ^6 M_{Pl}\left(7 (C^{(1)}_{H^2\varphi^2})^2+40 C^{(1)}_{H^2\varphi^2} C^{(2)}_{H^2\varphi^2}+60 (C^{(2)}_{H^2\varphi^2})^2\right)}{138915 \Lambda ^8},\\
F_5 &=-\frac{3 \sqrt{\frac{2}{5}} M_{Pl}\zeta (5)^2 \left(m_{\varphi}^2 \left(6 (C^{(1)}_{H^2\varphi^2})^2+5 C^{(1)}_{H^2\varphi^2} (9 C^{(2)}_{H^2\varphi^2}+2 d_3)+30 C^{(2)}_{H^2\varphi^2} (3 C^{(2)}_{H^2\varphi^2}+d_3)\right)\right)}{5 \pi ^6 \Lambda ^8}\nonumber\\
&\, \, \, \, -\frac{3 \sqrt{\frac{2}{5}} M_{Pl}\zeta (5)^2 \left(10 m^2_H d_4 (C^{(1)}_{H^2\varphi^2}+3 C^{(2)}_{H^2\varphi^2})\right)}{5 \pi ^6 \Lambda ^8},\\
F_3 &= \frac{2\pi ^2 M_{Pl}\left(10 m_{\varphi}^2 (-4 c_3 (C^{(1)}_{H^2\varphi^2}+3 C^{(2)}_{H^2\varphi^2})+3 m^2_H C^{(1)}_{H^2\varphi^2} d_4+2m^2_H d_4 (6 C^{(2)}_{H^2\varphi^2}+d_3))\right)}{388800 \sqrt{10} \Lambda ^8}\nonumber\\
&\, \, \, \,+ \frac{\pi ^2 M_{Pl}\left(m_{\varphi}^4 \left(9 (C^{(1)}_{H^2\varphi^2})^2+20 C^{(1)}_{H^2\varphi^2} (3 C^{(2)}_{H^2\varphi^2}+d_3)+20 (3 C^{(2)}_{H^2\varphi^2}+d_3)^2\right)+20 m^4_H d_4^2\right)}{388800 \sqrt{10} \Lambda ^8},\\
F_1 &= \frac{\sqrt{\frac{5}{2}} M_{Pl}m_{\varphi}^2m^2_H  \zeta (3)^2 (-2c_3+d_4) \left(m^2_H d_4-m_{\varphi}^2 (C^{(1)}_{H^2\varphi^2}+6 C^{(2)}_{H^2\varphi^2}+2 d_3)\right)}{48 \pi ^6 \Lambda ^8},\\
F_{-1} &= -\frac{\sqrt{\frac{5}{2}} M_{Pl}m_{\varphi}^4 m^4_H (d_4-2 c_3)^2}{6912 \pi ^2 \Lambda ^8}.
\end{align}

As a result, in the limit of $T_\mathrm{reh} \gg m_H$, we can approximate the dark matter yield in Eq.~\eqref{eq:yield} to
\begin{align}
Y(m_H) \simeq Y(T_\mathrm{reh}) + \frac{g_\phi}{\sqrt{g_*(T_{\rm reh})}}
\left[F_7 T^7_\mathrm{reh} + F_5 T^5_\mathrm{reh} + F_3 T^3_\mathrm{reh} + F_1T_\mathrm{reh} - F_{-1}\frac{1}{m_H}\right].
\end{align}
Furthermore, assuming $Y_\varphi(T_\mathrm{reh}) = 0$ and $d_4, c_3={\cal O}(1)$, from $T_\mathrm{reh} \gg m_\varphi, m_H$, we can make a further approximation for the dark matter yield by
\begin{align}
Y_\varphi(m_H) \simeq \frac{g_\phi}{\sqrt{g_*(T_{\rm reh})}} F_7 T^7_\mathrm{reh}. 
\label{eq:yield2}
\end{align}
Then, the dark matter production occurs at the high temperature near the reheating temperature, so we can fix $g_*= 106.75$ in the SM.

Finally, the dark matter relic abundance at present is determined as
\begin{align}
\Omega_\varphi h^2 &= 1.6\times 10^8 \left(\frac{m_\varphi}{1~\mathrm{GeV}}\right)\left(\frac{g_{*s}(T_0)}{g_{*s}(T_{\rm reh})}\right) Y_\varphi(m_H)\\
&\simeq 5.7\times 10^5 \left(\frac{m_\varphi}{1~\mathrm{GeV}}\right)
 \frac{8 \sqrt{\frac{2}{5}} \pi ^6 M_{Pl}T^7_\mathrm{reh}  \left(7 (C^{(1)}_{H^2\varphi^2})^2+40 C^{(1)}_{H^2\varphi^2} C^{(2)}_{H^2\varphi^2}+60 (C^{(2)}_{H^2\varphi^2})^2\right)}{138915 \Lambda ^8}.
\label{eq:DM_relic}
\end{align}
Here, the number of the effective entropy degrees of freedom is given by $g_{*s}(T_0) = 3.91$ at present and $g_{*s}(T_{\rm reh})=106.75$ at the reheating temperature.

\section*{Acknowledgments}

The work is supported in part by Basic Science Research Program through the National Research Foundation of Korea (NRF) funded by the Ministry of Education, Science and Technology (NRF-2022R1A2C2003567 and NRF-2021R1A4A2001897). 


\end{document}